\documentclass{PoS}
\usepackage{graphicx}

\title{Event-plane dependent away-side jet-like correlation shape in Au+Au collisions at $\sqrt{s_{NN}}=200$ GeV from STAR}

\ShortTitle{EP-dependent away-side jet-like correlations in STAR}

\author{\speaker{Liang Zhang (for the STAR collaboration)}\\
        Key Laboratory of Quark and Lepton Physics (MOE) and Institute of Particle Physics,\\
Central China Normal University, Wuhan 430079, China\\

 Department of Physics and Astronomy, Purdue University, West Lafayette, Indiana 47907, USA
        E-mail: \email{l.zhang@mails.ccnu.edu.cn}}

\abstract{We employ a data-driven method to subtract the flow background of all harmonics by calculating the difference of the two-particle correlations between the close-region and far-region, determined depending on the pseudo-rapidity ($\eta$) distance from the region where an enhanced recoil transverse momentum ($P_x$) from a high-$p_T$ trigger particle is selected. We analyze the correlation shape as a function of the trigger particle azimuthal angle relative to the event-plane (EP) reconstructed from the beam-beam counters (BBCs) which are displaced by several units in $\eta$ from the mid-rapidity region. The large $\eta$ gap can effectively eliminate the auto-correlation between trigger particles and EP. We correct for the relatively large resolution effect from the BBC EP determination via an unfolding procedure. 
The width of unfolded away-side jet-like correlation increases with longer path-length, which is an indication of jet-medium interactions.

}

\FullConference{International Conference on Hard and Electromagnetic Probes of High-Energy Nuclear Collisions\\
		30 September - 5 October 2018\\
		Aix-Les-Bains, Savoie, France}

\begin{document}
\section{Introduction}
A strongly coupled quark gluon plasma (QGP) is believed to be created in relativistic heavy-ion collisions~\cite{Adams:2005dq}.
Jet-like correlations are a good probe of the energy loss mechanism of hard partons traversing the QGP medium~\cite{Jacobs:2004qv,  Wang:2013qca, Connors:2017ptx}. They are often analyzed by calculating the azimuthal angle difference ($\Delta\phi$) between high transverse momentum ($p_{T}$) trigger particles and associated particles.
While the near-side ($|\Delta\phi|<\pi/2$) correlations (in the trigger particle hemisphere) are not much modified, indicating surface bias of these correlations~\cite{Jacobs:2004qv}, the away-side ($|\Delta\phi-\pi|<\pi/2$) correlations recoiling from the trigger particles are significantly modified: suppressed at high $p_T$ and broadened at low $p_T$~\cite{Wang:2013qca, Connors:2017ptx, Adams:2005ph}. For non-central Au+Au collisions, the in-medium path length that the recoil (away-side) parton traverses is expected to depend on its emission angle with respect to the reaction plane (RP)~\cite{Agakishiev:2010ur, Agakishiev:2014ada}, spanned by the impact parameter and beam directions and which is approximated by the final state event plane (EP). 
In these proceedings, we investigate the EP dependence of the away-side jet-like correlation shape. 
\section{Analysis Method}
Measurements of jet-like correlations in heavy-ion collisions are complicated by the large underlying background~\cite{Wang:2013qca}. 
A novel method to subtract all harmonic flow backgrounds without assumptions on their amplitude and shape~\cite{KunProceedings} is used in this analysis.
We first select events with a large recoil transverse momentum ($P_x$) to a high-$p_T$ trigger particle to enhance the away-side jet population for a specific forward or backward pseudo-rapidity ($\eta$) region ($-1<\eta<-0.5$ or $0.5<\eta<1$). $P_x$ is given by
\begin{eqnarray}\label{eq:Px}
&P_{x}|^{\eta_2}_{\eta_1}= &\sum_{\eta_1 < \eta < \eta_2, |\phi-\phi_{trig}| > \pi/2} p_T \cos(\phi-\phi_{trig})\frac{1}{\epsilon},\nonumber
\end{eqnarray}
where all charged particles ($0.15<p_T<10$ GeV/$c$) in the opposite hemisphere of the trigger particle within a given $\eta$ range are included.  We use the inverse of single-particle tracking efficiency ($\epsilon$) to correct for particle detection efficiency.
Then two $\eta$ regions ($-0.5<\eta<0$ and $0<\eta<0.5$) are defined as the close-region and far-region, respectively, depending on the distance to the $\eta$ region where the $P_x$ is calculated.
We analyze the two-particle correlations between the trigger and associated particles in the close-region and far-region separately.
The anisotropic flow contributions to these two regions are nearly equal because these two regions are symmetric about mid-rapidity.
Therefore, the flow contributions to the close-region and far-region are cancelled out in the correlation difference.
The away-side jet contribution to the close-region should be significantly larger than that to the far-region because of the different $\eta$ distances.
The difference between the close- and far-region two-particle correlations, therefore, contains predominantly the contribution from away-side jet-like correlations, hence is a good measure of the correlation shape. 

The $2^{nd}$ order harmonic EP~\cite{Voloshin:1994mz} is reconstructed with the beam-beam counters (BBCs).
The $\eta$ ranges of the BBCs are $3.3<|\eta|<5.2$.
The trigger and associated particles are detected by the Time Projection Chamber (TPC) at mid-rapidity ($|\eta|<1$).
The large $\eta$ gap between the TPC and BBCs can effectively eliminate the auto-correlation between trigger particles and EP.
The resolution of the reconstructed EP from the BBCs is calculated with the two sub-event method~\cite{Voloshin:1994mz}, and is found to be 0.135$~\pm~$0.002 (stat.) in 20-60\% Au+Au collisions at $\sqrt{s_{NN}}=200$~GeV.
This is a measure of its accuracy in representing the true EP, and is relatively poor. Future measurement by STAR's recently installed Event Plane Detector will improve the EP resolution.
\section{Results}
\begin{figure}[ht!]
\centering
\includegraphics[width=\textwidth]{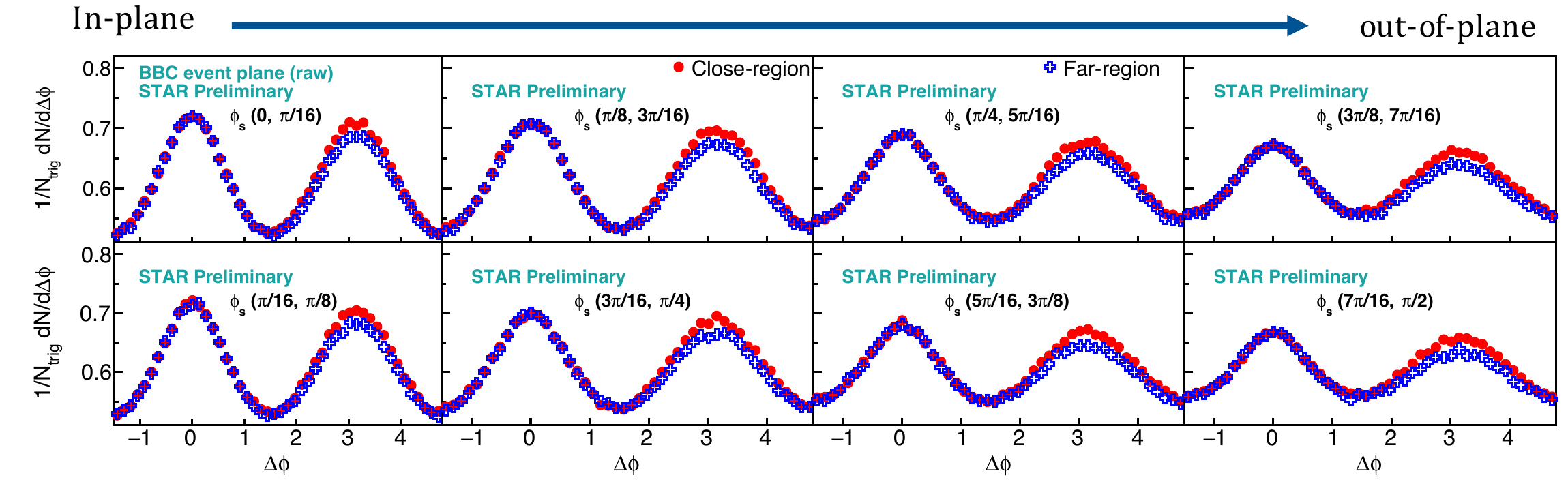}
\caption{Two-particle azimuthal correlations in the close-region (red solid circles) and far-region (blue open crosses) for different $\phi_{s}$ bins for 3$~<p_{T}^{trig}<~$10~GeV/$c$ and 1$~<p_{T}^{assoc}<~$2~GeV/$c$ in 20-60\% Au+Au collisions at $\sqrt{s_{NN}}=200$~GeV.}
\label{fig:raw_correlations}   
\end{figure}
\begin{figure}[ht!]
\centering
\includegraphics[width=\textwidth]{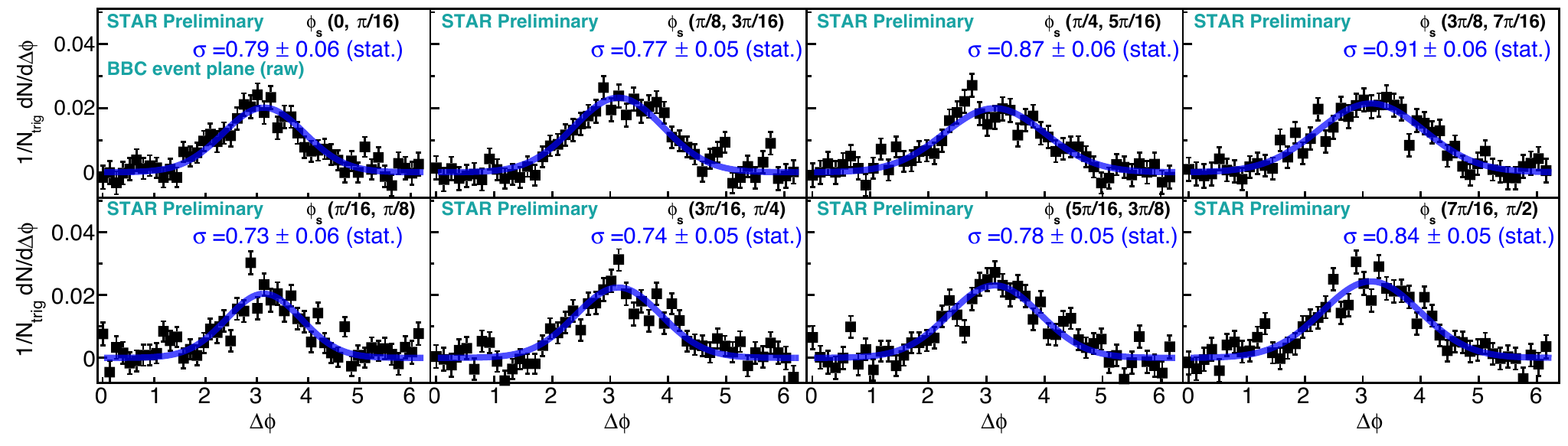}
\caption{The differences between the close-region and far-region two-particle correlations in Fig.~\ref{fig:raw_correlations}. Errors are statistical only. The blue curves are Gaussian fits with the mean value fixed at $\pi$.}
\label{fig:raw_differences}   
\end{figure}

Figure~\ref{fig:raw_correlations} shows the close- and far-region two-particle correlations in eight different $\phi_s$ bins with the trigger and associated particle $p_T$ ranges of 3$~<p_{T}^{trig}<~$10~GeV/$c$ and 1$~<p_{T}^{assoc}<~$2~GeV/$c$ in 20-60\% Au+Au collisions at $\sqrt{s_{NN}}=200$~GeV. Here $\phi_s$ is the trigger particle azimuthal angle relative to the reconstructed EP. The near-side correlations are well consistent in all $\phi_s$ bins between the close- and far-region. The ratios of the far- to close-region on the near side are approximately unity, with deviations less than 0.5\% (within 2$\sigma$ statistical uncertainty). 
This remaining deviation is normalized out before taking the correlation difference between the close-region and far-region, shown in Fig.~\ref{fig:raw_differences}.
The away-side correlations are different presumably due to away-side jet-like contributions.
We use a Gaussian function (with centroid fixed at $\pi$) to fit the differences in Fig.~\ref{fig:raw_differences} to extract the correlation widths. The fits are superimposed as the blue curves.
The Gaussian width ($\sigma$) increases modestly with $\phi_s$. \\

\begin{figure}[ht]
\centering
\includegraphics[width=\textwidth]{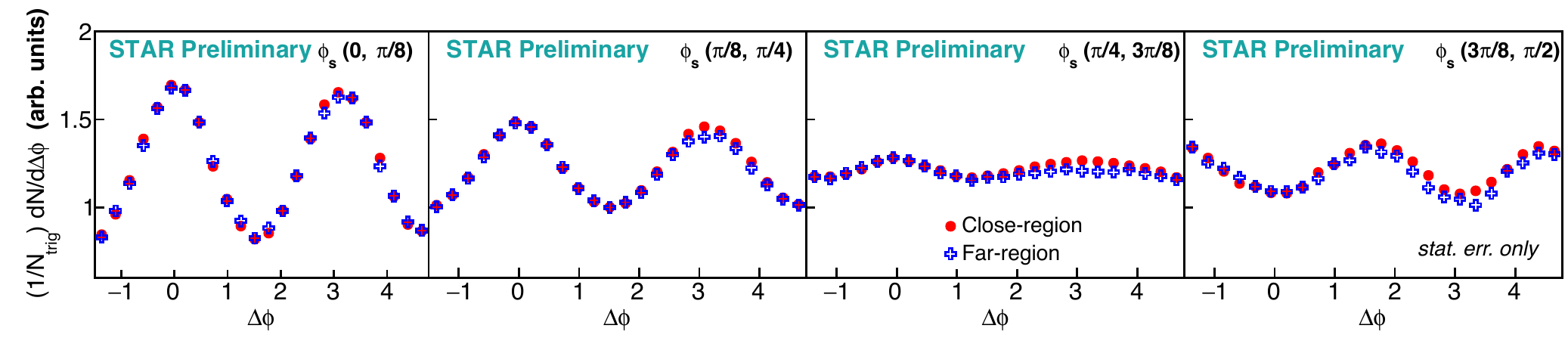}
\caption{The unfolded two-particle correlations in the close-region (red solid circles) and far-region (blue open crosses) from those in Fig.~\ref{fig:raw_correlations}.}
\label{fig:unfolded_correlations}   
\end{figure}

The away-side correlations in different $\phi_s$ bins are smeared significantly because of the poor EP resolution. We correct for this smearing effect by an unfolding procedure as follows.
We take the measured trigger particle distribution in $\phi_s$ and the EP resolution as inputs.
The true $\phi_s$ distribution is obtained by amplifying the Fourier modulation of the measured $\phi_s$ distribution by the inverse of the EP resolution factor~\cite{Voloshin:1994mz}.
Similarly, the distribution of azimuthal angle difference between the measured EP and true EP is evaluated by the EP resolution~\cite{Voloshin:1994mz}.
The probability matrix ($\mathbf{A}$) is determined using Monte Carlo simulations, where the element $A_{ij}$ is the probability for the measured $\phi_s$ in the $j^{th}$ bin to come from the true $\phi_s$ in the $i^{th}$ bin.
For each $\Delta\phi$ bin, we take the eight amplitudes of the two-particle correlations in eight $\phi_s$ bins (as shown in Fig.~\ref{fig:raw_correlations}) as the input in the unfolding procedure.
We use a least-squares method with Tikhonov regularization~\cite{Regular_a} as implemented in the TUnfold package~\cite{Tunfold}.
The best value of the regularization strength ($\tau^{2}$) is obtained via implementing the L-curve scan in TUnfoldDensity.
We set the number of unfolded bins to be half of the input in our analysis. 
We repeat the unfolding procedure for all $\Delta\phi$ bins and obtain the unfolded correlation results.
Figure~\ref{fig:unfolded_correlations} shows the unfolded two-particle correlations in four $\phi_s$ bins. The $\Delta\phi$ bins are rebinned by two to reduce the point-to-point fluctuations. It is found that the unfolded correlation shape in the out-of-plane ($3\pi/8<\phi_{s}<\pi/2$) direction is significantly  different from the measured correlation shape.
This is a result of the poor EP resolution.

\begin{figure}[ht]
\centering
\includegraphics[width=\textwidth]{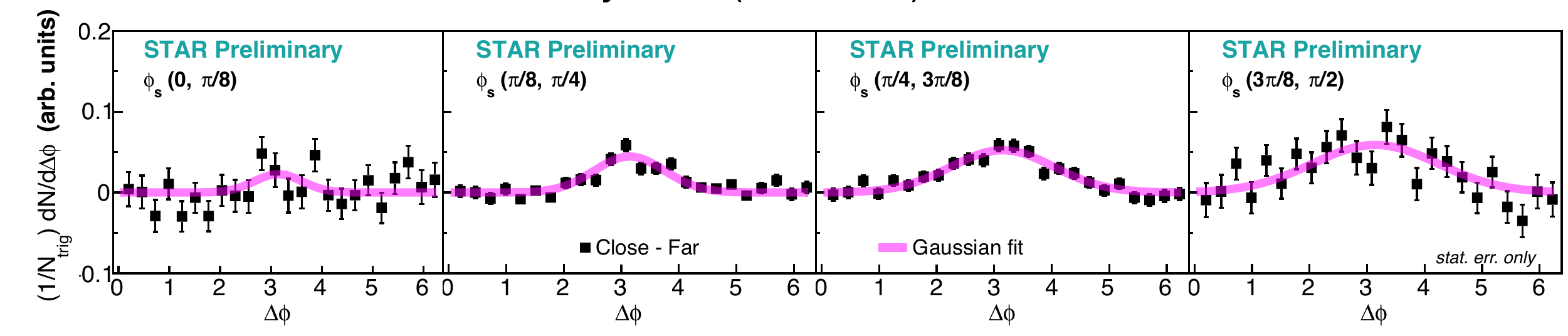}
\caption{The differences between the unfolded close-region and far-region two-particle correlations in Fig.~\ref{fig:unfolded_correlations}. Errors are statistical only. The pink curves are Gaussian fits with the mean value fixed at $\pi$.}
\label{fig:unfolded_differences}   
\end{figure}

Figure~\ref{fig:unfolded_differences} shows the differences between the unfolded close- and far-region two-particle correlations. The most in-plane and out-of-plane results have greater uncertainties after unfolding.
We also use a Gaussian function to fit the data points to obtain the correlation width. The fits are superimposed as the pink curves.\\

\begin{figure}[h]
\centering
\includegraphics[height=5cm]{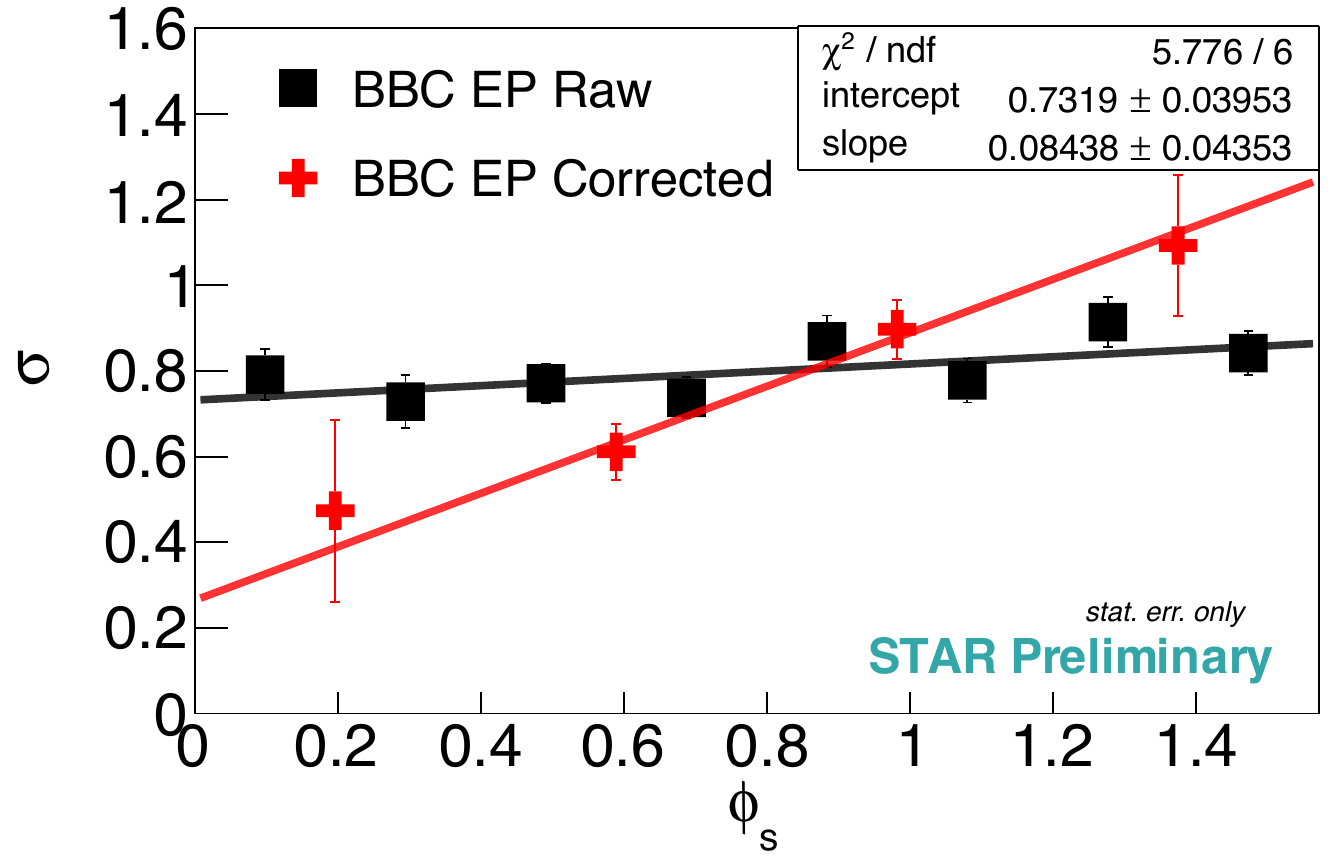}
\caption{The raw (black squares) and unfolded (red crosses) away-side correlation widths ($\sigma$) as a function of $\phi_{s}$. The black and red lines are corresponding linear fits.}
\label{fig:unfolded_sigma_vs_phis} 
\end{figure}

Figure~\ref{fig:unfolded_sigma_vs_phis} shows the comparison between the raw and unfolded away-side correlation widths as a functions of $\phi_s$.
The black and red lines are linear fits to the widths.
The slopes of the raw and unfolded results are 0.08$~\pm~$0.04 (stat.) and 0.66$~\pm~$0.27 (stat.) respectively.
Because the errors on the widths of the unfolded correlations are correlated among the $\phi_s$ bins, we estimate the statistical error on the unfolded slope as follows: (1) we randomly vary the data points in Fig.~\ref{fig:raw_correlations} using Gaussian sampling according to their statistical errors; (2) we use the same procedure to unfold the varied data points and extract a new Gaussian width after unfolding; (3) we obtain the linear slope of the new Gaussian width as a function of $\phi_s$;
and (4) we repeat step (1) - (3) many times to obtain a distribution of the slope and take the Gaussian width of the distribution as the statistical uncertainty on the slope. 
As seen from Fig~\ref{fig:unfolded_sigma_vs_phis}, the  unfolded away-side jet-like correlation width increases with $\phi_s$, providing a hint of jet-medium interactions.

\section{Summary}
We have applied a data-driven method to subtract flow backgrounds of all harmonics in jet-like correlations relative to high-$p_T$ trigger particles ($3<p_{T}^{trig}<10$ GeV/$c$) in Au+Au collisions at $\sqrt{s_{NN}}=200$~GeV.
The event-plane dependence of the away-side jet-like correlation shape is reported.
The $2^{nd}$ order EP is reconstructed with BBCs and the EP resolution is corrected via an unfolding procedure.
The Gaussian width of the away-side jet-like correlation is found to increase with $\phi_{s}$, providing a hint of jet-medium interactions.

\end{document}